\documentclass[reprint,amsmath,amssymb,aps,prl,superscriptaddress,]{revtex4-2}

\usepackage{subfiles}
\usepackage{graphicx}
\usepackage{dcolumn}
\usepackage{bm}
\usepackage{color}
\usepackage{braket}
\usepackage{hyperref}
\hypersetup{hidelinks}

\begin{document}

\preprint{APS/123-QED}

\title{Frequency-stable nanophotonic microcavities via integrated thermometry}



\affiliation{Department of Applied Physics and Applied Mathematics, Columbia University, New York, NY 10027, United States}
\affiliation{Department of Applied Physics and Ginzton Laboratory, Stanford University, Stanford, CA 94305, United States}
\affiliation{Department of Electrical Engineering, Columbia University, New York, NY 10027, United States}

\author{Sai Kanth Dacha}
\affiliation{Department of Applied Physics and Applied Mathematics, Columbia University, New York, NY 10027, United States}
\author{Yun Zhao}
\affiliation{Department of Applied Physics and Applied Mathematics, Columbia University, New York, NY 10027, United States}
\affiliation{Department of Applied Physics and Ginzton Laboratory, Stanford University, Stanford, CA 94305, United States}
\author{Karl J. McNulty}
\affiliation{Department of Electrical Engineering, Columbia University, New York, NY 10027, United States}
\author{Gaurang R. Bhatt}
\affiliation{Department of Electrical Engineering, Columbia University, New York, NY 10027, United States}
\author{Michal Lipson}
\affiliation{Department of Electrical Engineering, Columbia University, New York, NY 10027, United States}
\affiliation{Department of Applied Physics and Applied Mathematics, Columbia University, New York, NY 10027, United States}
\author{Alexander L. Gaeta}
\thanks{a.gaeta@columbia.edu}
\affiliation{Department of Applied Physics and Applied Mathematics, Columbia University, New York, NY 10027, United States}
\affiliation{Department of Electrical Engineering, Columbia University, New York, NY 10027, United States}

\date{\today}

\begin{abstract}
Field-deployable integrated photonic devices co-packaged with electronics will enable important applications such as optical interconnects, quantum information processing, precision measurements, spectroscopy, and microwave generation. Significant progress has been made over the past two decades on increasing the functional complexity of photonic chips. However, a critical challenge that remains is the lack of scalable techniques to overcome thermal perturbations arising from the environment and co-packaged electronics. Here, we demonstrate a fully integrated scheme to monitor and stabilize the temperature of a high-$Q$ microresonator on a Si-based chip, which can serve as a photonic frequency reference. Our approach relies on a thin-film metallic resistor placed directly above the microcavity, acting as an integrated resistance thermometer, enabling unique mapping of the cavity's absolute resonance wavelength to the thermometer's electrical resistance. Following a one-time calibration, the microresonator can be accurately and repeatably tuned to any desired absolute resonance wavelength using thermometry alone with a root-mean squared wavelength error of $<0.8$~pm over a timespan of days. We frequency-lock a distributed feedback (DFB) laser to the microresonator and demonstrate a 48× reduction in its frequency drift, resulting in its center wavelength staying within ±0.5 pm of the mean over the duration of 50 hours in the presence of significant ambient fluctuations, outperforming many commercial DFB and wavelength-locker-based laser systems. Finally, we stabilize a soliton mode-locked Kerr comb without the need for photodetection, paving the way for Kerr-comb-based photonic devices that can potentially operate in the desired mode-locked state indefinitely.
\end{abstract}

\maketitle


Optical frequency references are critically important tools for metrology~\cite{Hafele1972,Evenson1972,Hong_2017}, optical atomic clocks~\cite{LudlowBoydYe2015,Takano2016,Maurice:20}, precision spectroscopy~\cite{Udem2002,Hong:03,Thorpe:08}, and telecommunication systems~\cite{Nakagawa:96,Jiang:05,Roy2020}. At wavelengths away from molecular, atomic, and nuclear transitions, optical cavities with narrow linewidths and low long-term drift are indispensable for laser frequency stabilization. Cavities with large optical mode volumes and low Brownian thermorefractive noise (TRN) offer the best frequency stability, such as bulk single-crystal Silicon Fabry-Perot (FP) cavities with a reported fractional frequency instability of $10^{-17}$ over one second \cite{Kessler2012,Matei2017}. For many applications that require portability and robustness, such as in telecommunication systems, cryogenically-cooled bulk cavities are not well-suited, necessitating the use of compact reference and laser cavities that trade off size, weight, and power consumption (SWaP) for frequency stability that is typically several orders of magnitude worse compared to state-of-the-art bulk systems. In addition to the plethora of commercially available temperature-stabilized distributed feedback (DFB) and distributed Bragg reflector (DBR) lasers, several compact frequency stabilization technologies have been demonstrated in recent years, including integrated optical frequency discriminators~\cite{vanRees2023,Idjadi2024}, vapor-cell-integrated photonic chips~\cite{Hummon:18}, compact wavelength lockers~\cite{Suzuki2022}, and fiber interferometers~\cite{Kong2015}.

The rapid development of low-loss dielectric photonic platforms over the past two decades offers the prospect of realizing stable reference cavities in small form factors~\cite{Lee2013,Zhao:21}. In recent years, microcavities with large quality factors ($Q$s) have been shown to enable short-term frequency instability approaching $10^{-14}$~\cite{Guo2022,Heim2025}. However, achieving long-term stability remains challenging due to the enhanced sensitivity to fundamental and ambient temperature fluctuations~\cite{Huang2019,Panuski2020}. High-$Q$ microcavities are also excellent platforms for the generation of frequency combs, which have a number of applications such as spectroscopy~\cite{HanschNobel2006,Yu2018,Stern2020}, chip-scale multi-wavelength sources~\cite{Levy2010,Marin-Palomo2017}, microwave generation~\cite{Zhao2024,Sun2024,Kudelin2024}, light detection and ranging (LiDAR)~\cite{Kuse2019,Riemensberger2020}, and optical interconnects for large-scale machine learning~\cite{Rizzo2023,Wang2024}.

A central challenge that remains to be addressed in integrated photonics broadly is the lack of resource-efficient techniques to stabilize against thermal perturbations arising from ambient sources, co-packaged electronics, and thermal crosstalk~\cite{Herman2020,Orlandin2023}. For Kerr-comb applications, small temperature fluctuations can readily lead to pump detunings that lie outside of the modelocked regime of operation. Thermal crosstalk is a key bottleneck for neuromorphic computing applications~\cite{Orlandin2023} and densely integrated co-packaged photonic input/output (I/O) modules that require simultaneous tuning of multiple on-chip devices~\cite{Wang2024}.

Passive methods to mitigate thermal perturbations have been demonstrated previously, such as large mode volumes \cite{Lee2013}, and manipulating heat flow using undercuts \cite{Earnshaw:24} and trenches \cite{Gilardi:14}. However, passive methods only partially reduce the impact of thermal perturbations while adding significantly to fabrication cost and complexity. Active stabilization of silicon microresonators, based on the photoconductive effect, to a reference laser has been shown~\cite{Tria2024,Jayatilleka:15,Morichetti2014}. However, such a method is inherently limited by the reference laser's stability and is not applicable to high-$Q$ dielectric platforms that are commonly used for realizing Kerr combs. Laser frequency stabilization via dual-mode optical thermometry using a high-$Q$ microresonator has also been reported~\cite{Lim2019,Zhao:21}. However, it requires multiple phase modulators, cumbersome electronics, and the excitation of modes with proximate resonance frequencies and very high $Q$s. For comb-based devices, while active feedback using the optical output has been reported~\cite{Lin:19,Rebolledo-Salgado:24}, the requirement of additional drop-ports, optical filters, photodiodes, and free-space/fiber optics significantly adds to footprint and fabrication cost.

Here, we demonstrate a highly general approach, based on fully integrated thermometry, to stabilize the temperature of a high-$Q$ microcavity in the presence of significant external perturbations. A thin-film Platinum resistor is designed to exhibit a strong temperature-dependent electrical resistance and is used as an integrated resistance thermometer to measure a microcavity's temperature without the need for any photodetection or other integrated non-linear electronic elements (e.g., diodes, transistors). Similar metallic structures have previously been extensively used as microheaters~\cite{Miller:15,Joshi:16,Xue:16,Moille2022}. Platinum's chemical stability against basic chip cleaning reagents and atmospheric humidity ensures device longevity and long-term repeatability of our measurements. By engineering a dual-resistor scheme that achieves proper thermal stability of the microresonator waveguide core, we show that the absolute resonance frequency of an unshielded and unpackaged high-$Q$ silicon nitride (SiN) microcavity can be stabilized to within tens of MHz of any desired value and can be tuned repeatably by simply adjusting an electrical voltage. The frequency noise of such a microcavity is shown to be impervious to environmental drift and thermal crosstalk and limited only by $1/f$ flicker noise in the control electronics. We demonstrate that a one-time calibration of the resonance wavelength versus thermometer resistance is sufficient to repeatably tune the cavity to any desired resonance wavelength within the calibration range using thermometric control alone for at least a few days, removing the requirement for optical referencing or monitoring.

By locking a DFB laser to the stabilized microresonator, we demonstrate that it can serve as a fully-integrated photonic frequency reference for telecommunications transceiver applications. The locked laser's emission wavelength remains within $<\pm0.5$~pm of the mean over $50$~hours of measurement duration, outperforming many commercial DFB and wavelength-locker-based laser systems. Finally, we apply integrated thermometry to demonstrate a fully-thermally-stabilized Kerr comb that remains stably modelocked in the presence of strong thermal crosstalk. We envision this work to enable the realization of highly robust and field-deployable linear and nonlinear photonic devices for classical and quantum applications, and integrated Kerr combs that can potentially operate indefinitely.

\section{Results}

\begin{figure*}[!htbp]
\centering
\includegraphics[width=\textwidth]{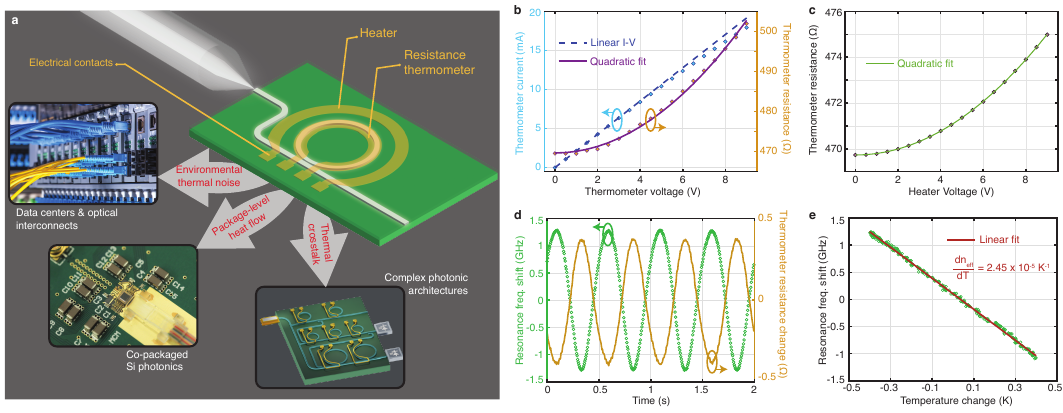}
\caption{\textbf{Integrated thermometry.} \textbf{a}, Conceptual illustration of integrated thermometry applied to stabilize a high-Q monolithic microresonator against thermal fluctuations arising from ambient heat sources and crosstalk from other thermally-tuned devices on the same chip. A thin-film metallic (Platinum) resistor, typically used as a microheater, exhibits a temperature-dependent resistance due to the metal’s natural temperature-dependent resistivity. As a result of the low heat capacity of the thin-film resistor, small fluctuations in the heat flow in its vicinity lead to large changes in its temperature, enabling the real-time monitoring of on-chip temperature by simply measuring its electrical resistance. A second identical resistor is used as a heater to perform active stabilization purely using the resistance thermometer, thereby eliminating the need for optical probing for stabilization. Image for data center sourced from Adobe Stock, Ref.~\cite{dataCent}. Image courtesy for picture of co-packaged device: Kaylx Jang. \textbf{b}, Current-voltage-resistance (I-V-R) characteristics of the resistance thermometer fabricated in this work. The I-V measurements (pale blue dots) deviate from the linear trend (dark blue line), and the resistance (yellow dots) exhibits a quadratic dependence (purple line) on the thermometer voltage. \textbf{c}, The measured thermometer resistance (lavender dots) also exhibits a quadratic dependence (green line) on the voltage applied across the (second) heater, indicating a linear dependence on power dissipated across the heater. \textbf{d}, Resonance frequency shift (green circles) of a high-Q microcavity of free-spectral range (FSR) 76 GHz, measured using a calibrated piezo-tuned probe laser, and measured change in thermometer resistance (gold line), for a sinusoidal perturbation applied via the heater. The cavity temperature and resonance frequency exhibit a strong anti-correlation. \textbf{e}, Measured resonance frequency shift plotted versus temperature measured by the thermometer, same data as in \textbf{d}. The linear fit provides an in-situ measurement of the thermo-optic coefficient of the fundamental spatial mode of the cavity waveguide.}\label{fig1}
\end{figure*}

Figure~\ref{fig1}(a) conceptually illustrates integrated thermometry. A thin-film metallic resistor with a temperature-dependent resistance placed directly above a high-$Q$ microcavity acts as an on-chip resistance thermometer. Due to the low heat capacity of the thin-film resistor, small heat fluxes translate to large and observable changes in its temperature. The current-voltage-resistance (I-V-R) characteristics displayed in Fig.~\ref{fig1}(b) show a non-linear I-V relationship, and a quadratic dependence of R on V, indicating a linear dependence of R on the dissipated power. This behavior is similar to that of a filament lamp, which exhibits lower (higher) resistance at lower (higher) voltages due to lower (higher) filament temperature from Joule heating. Figure~\ref{fig1}(c) shows that the thermometer resistance also depends quadratically on the voltage applied across the second heater, demonstrating that temperature changes induced by external heat sources can be measured by simply monitoring the resistance. Extended Data Fig.~\ref{extDataFig:TCR} shows an independent measurement of the temperature-dependent resistance~\cite{Bhatt2020}.

We combine the temperature measurement scheme described above with a scanning-laser optical probe to perform rapid in-situ measurements of the thermo-optic coefficient of a silicon-nitride (SiN) microresonator's waveguide mode. The free-spectral range (FSR) is 76 GHz, and the loaded $Q$ is approximately $3\times10^{6}$. The measurement procedure is described in Methods. Figure~\ref{fig1}(d) shows the measured resonance frequency shift and the corresponding change in thermometer resistance, illustrating a strong correlation. Figure~\ref{fig1}(e) plots the resonance frequency shift versus the measured temperature change. The slope of the linear fit is used to extract the thermo-optic coefficient of the fundamental spatial mode of the SiN waveguide, $dn_{\mathrm{eff}}/dT=2.45\times 10^{-5}$~K\textsuperscript{-1}. This value is approximately equal to the thermo-optic coefficient $dn/dT$ of SiN due to the high spatial confinement offered by the 2100-nm waveguide and is consistent with values measured using other techniques~\cite{Arbabi:13,Elshaari2016,Nejadriahi:20}.

\begin{figure*}[!htbp]
\centering
\includegraphics[width=\textwidth]{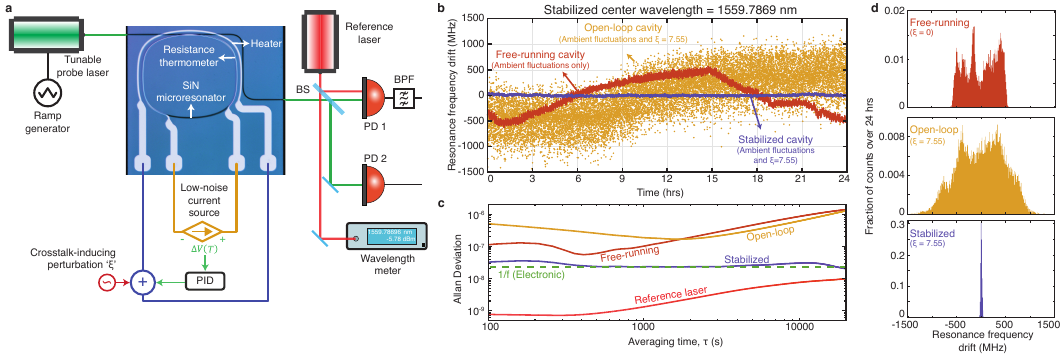}
\caption{\textbf{Long-term stabilization of the absolute resonance frequency of a high-Q SiN microresonator using integrated thermometry. a}, Simplified experimental schematic, depicting a tunable probe laser that is piezo-scanned across a microcavity resonance. Absolute frequency calibration of the scan is performed by producing a heterodyne note between the tunable laser and a stable reference laser, and simultaneously tracking the reference laser’s wavelength drift using a benchtop wavelength meter. The heterodyne beat note as well as the cavity resonance line-shape are measured synchronously with the piezo sweep using a real-time oscilloscope connected to a computer interface for periodic data acquisition every 10 seconds. For $\xi$=1, the pseudorandom perturbation applied via the heater yields a root mean-squared (RMS) of the induced cavity resonance frequency fluctuation that is equal to the cavity resonance full-width-at-half-maximum (FWHM) linewidth, which is 75 MHz for the microcavity fabricated in this work. \textbf{b}, Drift in the microcavity’s calibrated resonance frequency over the duration of 24 hours, for free-running ($\xi$=0), open-loop ($\xi$=7.55), and stabilized ($\xi$=7.55) operation. The free-running cavity exhibits a slow but strong drift in its resonance frequency due to ambient temperature drift in the lab, whereas the open-loop cavity additionally exhibits strong ``fast" fluctuations due to the perturbation introduced. The stabilized cavity’s resonance frequency remains highly stable in the presence of very strong ambient and crosstalk-induced perturbations. \textbf{c}, Allan Deviation (ADEV) versus averaging time, computed using the data displayed in b. The stabilized cavity exhibits an ADEV that is well below the free-running and open-loop modes of operation. The shape of the ADEV curve of the stabilized cavity matches that of $1/f$ noise, indicating that performance is only limited by control electronics. \textbf{d}, Histogram plots depicting the fraction of time over 24 hours (y-axis) that the microcavity’s absolute resonance frequency drift lies within a given frequency bin with a width of 5 MHz.}\label{fig2}
\end{figure*}

Figure~\ref{fig2}(a) illustrates the experimental schematic for investigating the long-term frequency stability of the microresonator using a piezo-controlled scanning probe laser. The absolute frequency calibration procedure is described in Methods. The microcavity resonance frequency is averaged over and recorded once every 10 seconds over the period of 24~hours. Three such measurements are performed, each under the following conditions, (i) `free-running': the cavity is subject only to ambient temperature fluctuations, (ii) `open-loop': additional perturbations are induced via a voltage-summing circuit connected to the heater to mimic thermal crosstalk from co-packaged electronics, and (iii) `stabilized': the microresonator's temperature is actively stabilized by applying proportional-integral-derivative (PID) feedback to the heater based on the voltage drop across the resistance thermometer connected to a low-noise current source. The strength of the perturbation applied is denoted by the empirical parameter $\xi$: for $\xi$=1, the root mean-squared (RMS) fluctuation of the resonance frequency equals the short-term resonance full-width-at-half-maximum (FWHM) linewidth, which is 75 MHz for the microcavity used in this work. A circuit schematic of the current source and its noise performance are provided in Supplementary Fig.~S1. The thermal design of the dual-resistor scheme is explained in Extended Data Fig.~\ref{extDataFig:thermDesign}.

%

Figure~\ref{fig2}(b) displays the measured resonance frequency fluctuations. Without stabilization, ambient noise causes a significant resonance drift, even in a temperature-stabilized laboratory, as seen in the `free-running cavity' trace. The resulting standard deviation of the resonance frequency is $\approx341$~MHz. The additional perturbation applied ($\xi=7.55$) drives faster-timescale fluctuations seen in the `open-loop cavity' trace. The measured open-loop standard deviation is $\approx470$~MHz. Supplementary Figs.~S2 and S3 provide the temperatures measured by the resistance thermometer for the free-running and open-loop cases, which show strong correlations with the resonance frequency shifts. Finally, the stabilized cavity exhibits a highly stable resonance frequency with significantly weaker fluctuations (standard deviation $\approx13.7$~MHz, a $34\times$ reduction from the open-loop scenario) even in the presence of additional perturbation ($\xi=7.55$). Figure~\ref{fig2}(d) visualizes the time-domain data as histograms (bin size $=5$~MHz). Notably, even in the free-running case, the resonance frequency histogram does not exhibit a well-defined peak.

Figure~\ref{fig2}(c) shows the Allan deviation (ADEV) versus averaging time computed from the time-domain data. The stabilized cavity offers a significantly lower ADEV compared to the free-running and open-loop cases. The flat shape of the stabilized cavity's ADEV closely matches that of pure $1/f$ noise, indicating that unbounded random-walk resonance frequency drifts arising from environmental and crosstalk perturbations are effectively canceled out by the stabilization method, and that residual fluctuations are bounded and limited only by noise in the control electronics used in this work. Supplementary Fig.~S4 provides ADEV comparison measurements demonstrating that the stabilized microresonator offers superior long-term frequency stability compared to many commercially available telecommunication lasers~\cite{Wade:21,Wang2024} available in our laboratory, demonstrating its potential utility as a highly practical integrated reference cavity. 

\begin{figure*}[!htbp]
\centering
\includegraphics[width=\textwidth]{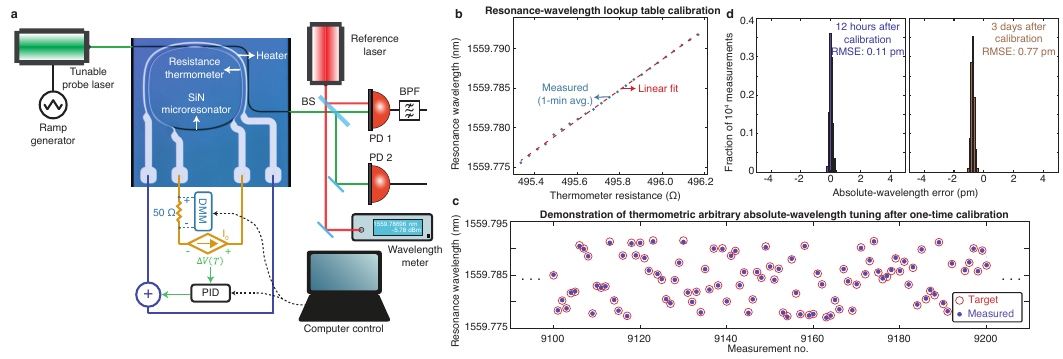}
\caption{\textbf{Demonstration of thermometric absolute-resonance-wavelength lookup table.} \textbf{a.}, Simplified experimental schematic, depicting a tunable probe laser that is piezo-scanned across a microcavity resonance. The absolute resonance wavelength is measured by producing a heterodyne beat between the tunable laser and a stable reference laser, and simultaneously tracking the reference laser’s wavelength using a highly-stable benchtop wavelength meter. The current produced by the low-noise current source is monitored by measuring the voltage across a series $50.0$ $\Omega$ bulk resistor using a precision 6.5-digit digital multimeter (DMM). The PID loop is implemented using a commercial FPGA, and the set-point voltage is input via a computer interface. \textbf{b.} Calibration measurement of the absolute resonance wavelength vs the thermometer resistance. The linear fit extracted from this one-time calibration is utilized in subsequent measurements to compute the desired thermometer resistance to achieve any desired resonance wavelength within the calibration range. \textbf{c.} Demonstration of arbitrary absolute-wavelength tuning of the microresonator via thermometry. A string of 10,000 pseudorandom ``target" wavelengths (red hollow circles) are provided as inputs to the thermometric control system illustrated in \textbf{a.},  which computes the desired temperature for each target wavelength and stabilizes the microresonator to the computed temperature. The resonance wavelength is measured (blue-magenta filled circles) and compared with the target wavelengths. A zoomed-in portion of 100 consecutive measurements out of 10,000 is shown. \textbf{d.} Histograms showing the deviation between the target and measured absolute resonance wavelengths following a one-time calibration. The histogram bin size is 0.1 pm. The first set of 10,000 measurements, which concluded 12 hours after the calibration run, exhibits a root-mean squared absolute-wavelength error of 0.11~pm. The second set of measurements three days later shows a similar narrow distribution of absolute-wavelength error, but exhibits a systematic error, resulting in a 0.77~pm.}\label{fig3}
\end{figure*}

The long-term repeatability of the microcavity's tunable resonance wavelength is illustrated by the demonstration of a resonance-wavelength thermometric lookup table. A one-time calibration measurement of the absolute resonance wavelength versus the thermometer resistance is performed using the setup described in Fig.~\ref{fig3}\textbf{a}. The current source output is monitored by measuring the voltage across a series $50.0$~$\Omega$ bulk termination resistor using a precision 6.5-digit digital multimeter (DMM; Agilent 34401A). The thermometer voltage is measured by subtracting the voltage across the termination from the total voltage across the current source, which in stabilized operation equals the PID set-point voltage. The thermometer resistance calculated as the ratio of these two quantities also includes a small contribution from contact resistance of the spring-loaded electrical contact probes  (PicoProbe 40A-GS-150-P) employed in our experimental setup. The results of the one-time calibration are shown in Fig.~\ref{fig3}\textbf{b}. The linear fit extracted from this calibration is used in subsequent measurements to compute the desired thermometer resistance to achieve any desired resonance wavelength within the calibration range.

A computer-generated string of $10,000$ pseudorandom ``target" resonance wavelengths is fed into a thermometric control loop implemented using a field-programmable gate array (FPGA), which calculates the desired thermometer resistance corresponding to a target resonance wavelength using the calibration fit, and stabilizes the microresonator to the computed temperature. The resonance wavelength is then measured and compared with the target value. Each such ``measurement" takes approximately five seconds. Two separate sets of 10,000 measurements each are performed. A zoomed-in portion of the first set of 10,000 measurements, which concluded 12 hours after the calibration and is displayed in Fig.~\ref{fig3}\textbf{c}, shows close agreement between the measured and target resonance wavelengths. Figure~\ref{fig3}(d) shows histograms of the absolute-wavelength error (bin size $=0.1$~pm), defined as the target minus measured wavelengths. For the first set, a root-mean squared error (RMSE) is $0.11$~pm is observed. A systematic error, resulting in a larger RMSE of $0.77$~pm, was observed in the second set carried out 3 days after the calibration. It is hypothesized to be stemming from drift in the contact resistance of the spring-loaded probes.

\begin{figure*}[!htbp]
\centering
\includegraphics[width=\textwidth]{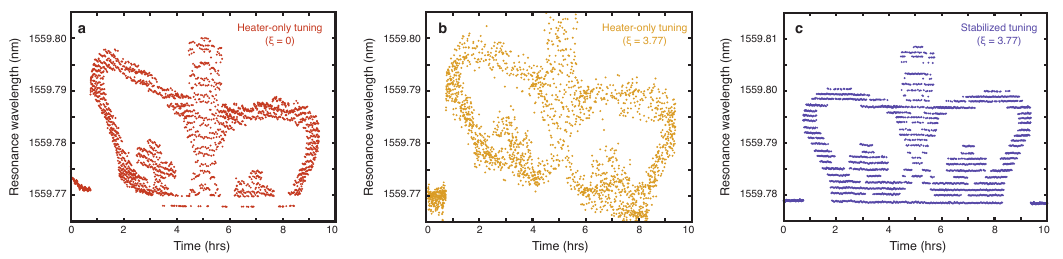}
\caption{\textbf{Tunability and long-term repeatability of the microresonator’s absolute resonance wavelength.} The resonance frequency is programmed to trace a pre-defined shape by blind tuning of the power supplied to the outer heater in the presence of ambient noise alone ($\xi$=0; shown in \textbf{a}), in the presence of additional perturbation ($\xi$=3.77; shown in \textbf{b}), and by tuning of the set-point for temperature stabilization in the presence of ambient noise as well as additional perturbation ($\xi$=3.77; shown in \textbf{c}). The measurements for heater-only tuning show that the microresonator’s resonance wavelength exhibits a large drift over the timescale of only a few hours, even for the same pre-programmed heater voltage. The stabilized tuning, on the other hand, yields a highly repeatable performance even in the presence of strong ambient and external noise.}\label{fig4}
\end{figure*}

The repeatability of the thermometrically controlled microcavity's resonance wavelength is further demonstrated by programming the resonance wavelength to trace a predetermined pattern. When the heater is actuated without feedback control, in the presence ($\xi=3.77$) as well as absence ($\xi=0$) of additional perturbation, significant long-term drift is observed over the $10$-hour measurement duration (Figs.~\ref{fig4}(a) and (b)). On the other hand, when the microresonator is stabilized and the set-point voltage is tuned instead, long-term drift is canceled out, and the resonance wavelength exhibits a highly repeatable relationship with the set-point voltage, as shown in Fig.~\ref{fig4}(c).

\begin{figure*}[!htbp]
\centering
\includegraphics[width=\textwidth]{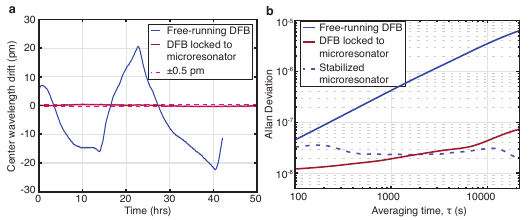}
\caption{\textbf{Demonstration of long-term laser frequency stabilization to the microresonator.} A DFB laser is locked to the resonance of the stabilized microresonator using the Pound-Drever-Hall (PDH) method, and its emission center wavelength is continually measured once every 10 seconds using a bench-top wavelength meter. \textbf{a}, Time-domain comparison of center wavelength drift of free-running and PDH-locked configurations of a DFB laser. The locked laser's center wavelength remains within a $\pm0.5$~pm window of the mean, indicated by the dashed lines, for 50 hours. The free-running laser is only subject to ambient temperature drift in a temperature-stabilized laboratory environment. \textbf{b}, Comparison of measured Allan Deviation (ADEV) of the free-running and locked configurations of the DFB laser. The locked laser exhibits slightly lower ADEV than the stabilized microresonator at lower averaging times due to the strong integral gain of the PDH feedback servo. The slightly higher ADEV at larger averaging times is attributed to insufficient gain of the PDH feedback servo at very low frequencies.}\label{fig5}
\end{figure*}

The frequency stability of the microresonator is transferred to a fiber-coupled DFB laser using the Pound-Drever-Hall (PDH) method. For an integrated laser scheme, self-injection locking may be more suitable~\cite{Guo2022}. The DFB emission wavelength is recorded once every 10~seconds using a wavelength meter. As shown in Fig.~\ref{fig5}(a), the free-running DFB experiences significant wavelength drift due to ambient temperature fluctuations, with a standard deviation of $11.5$~pm (i.e., $1.43$~GHz) over $42$~hours of measurement time. In comparison, the locked DFB is highly stable, with a peak-to-peak variation in emission wavelength that $<\pm0.5$~pm of the mean for over $50$~hours, and a standard deviation of $0.24$~pm (i.e., $30$~MHz). The stability enhancement factor is $\approx48\times$. The long-term stability reported here is superior to many commercially available DFB and wavelength-locker-based laser systems. Figure~\ref{fig5}(b) shows that the ADEV of the locked DFB is significantly lower and is comparable to the ADEV of the stabilized microresonator displayed in Fig.~\ref{fig2}. Due to the large integral gain of the PDH servo, a lower ADEV is observed at smaller averaging times. The higher ADEV at larger averaging times is attributed to insufficient gain of the PDH servo at very low frequencies. 

\begin{figure*}[!htbp]
\centering
\includegraphics[width=\textwidth]{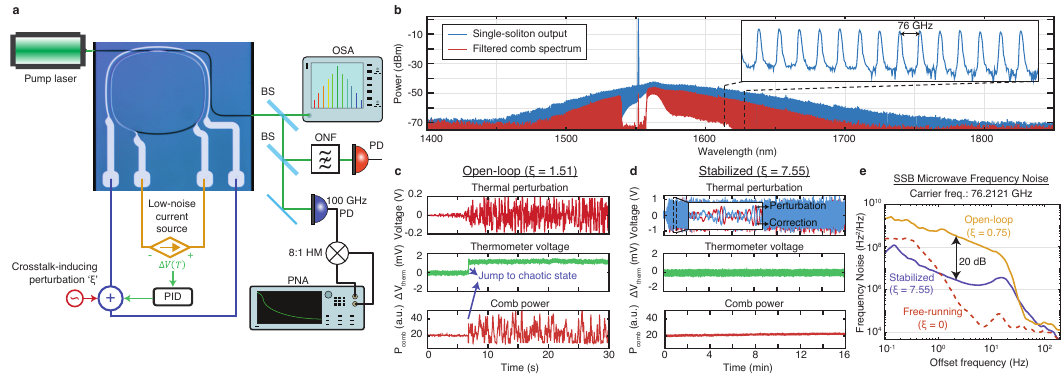}
\caption{\textbf{Fully-thermally stabilized Kerr optical frequency comb. a}, Simplified experimental schematic, depicting the comb generation and characterization setup. A soliton Kerr comb is generated in the microresonator by fast-tuning the current supplied to the resistance thermometer. The output comb is measured using an optical spectrum analyzer (OSA), and its frequency noise is measured using a high-speed photodetector connected to a phase-noise analyzer (PNA) via an 8:1 harmonic mixer (HM). The comb power is tracked using a slow photodiode (PD) after passing the output light through an optical notch filter (ONF). Comb stability in the presence of external noise is studied by applying a perturbation across the outer heater after generating the comb. \textbf{b}, Generated single-soliton and filtered output spectra, measured using an OSA. \textbf{c} and \textbf{d}, Perturbation voltage (top), measured temperature (middle), and output comb power (bottom), for open-loop and stabilization operations, for noise strengths of $\xi$=1.51 and $\xi$=7.55 respectively. Without stabilization, the comb jumps from a modelocked state to a chaotic regime almost instantly after the perturbation amplitude is increased. When stabilized, the comb remains modelocked even in the presence of much stronger noise. e, Measured single-sideband (SSB) frequency noise of the microwave generated by detecting the comb using a very-high-speed PD, for open-loop operation for a noise strength that is not weak enough to completely destabilize the comb ($\xi$=0.75; yellow line) and stabilized operation for a high noise strength ($\xi$=7.55; purple line). The microwave generated in the stabilized case has considerably lower noise than the open-loop case despite the much stronger noise.}\label{fig6}
\end{figure*}

Finally, we show in Fig.~\ref{fig6}(a) an approach to stabilize a modelocked Kerr comb, which is typically highly sensitive to thermal perturbations. A $1550$-nm pump laser is amplified before coupling into the photonic chip (pump power $=150$~mW). The thermometer current source is modulated to generate a single-soliton Kerr comb~\cite{Joshi:16}. The comb power is measured using a photodiode after filtering the pump line using a fiber-optic notch filter. The stability of the Kerr comb is studied by applying a white-noise perturbation (bandwidth $=20$~Hz) via the outer on-chip heater and measuring the on-chip temperature and the output comb power. The repetition-rate noise is measured by generating a microwave using a $100$-GHz bandwidth photodetector and characterizing it using a phase noise analyzer (PNA; Rohde \& Schwarz \textregistered~ FSWP50) via an 8:1 harmonic mixer.

For $\xi=1.51$ and without feedback correction, the modelocked Kerr comb transitions to a chaotic comb state almost immediately, as seen in Fig.~\ref{fig6}(c). The thermometer voltage plot shows a sudden increase, indicating an abrupt jump to a state with higher intracavity power, such as a chaotic comb. This is corroborated by the random fluctuations exhibited by the comb power and the thermometer voltage following the jump. The smaller amplitude of fluctuations observed by the thermometer following the jump result from an increase (decrease) in temperature at the thermometer due the perturbation and its consequent secondary effect, namely decrease (increase) in thermometer temperature due to decrease (increase) in intracavity optical power and self-heating. This is caused by the effectively blue-detuned pump of the chaotic comb and is not applicable to the modelocked comb due to its effectively red-detuned pump. 

In the stabilized case ($\xi=7.55$; Fig.~\ref{fig6}(d)), the PID correction voltage cancels the applied perturbation, leading to a stable temperature and comb power. Whereas the Kerr comb is pushed out of its existence range immediately in the open-loop ($\xi=1.51$) scenario, closed-loop operation was observed to support modelocking for over 20 minutes, limited only by the drift in the input fiber-to-chip optical coupling. In the absence of coupling drift, such as in packaged photonic devices, cold-cavity detuning can be made impervious to external thermal fluctuations. Thermometric control can then enable Kerr-comb-based photonic devices that potentially remain modelocked indefinitely.

Lastly, Fig.~\ref{fig6}(e) shows the measured microwave single-sideband (SSB) frequency noise spectra. For the `open-loop' scenario, the noise strength is reduced ($\xi=0.75$) to maintain modelocking. The microwave noise for the stabilized case is $>20$~dB below the open-loop case despite the stronger added-noise strength ($\xi=7.55$). At offset frequencies below $\sim1$~Hz, the stabilized cavity's microwave exhibits lower noise than the free-running microwave since ambient drift is more prominent on longer timescales. A comparison of the measured soliton repetition rates is presented in Supplementary Fig.~S5.




\section{Discussion}\label{sec6}
In summary, we have demonstrated a powerful new approach based on fully integrated thermometry to stabilize the resonance frequency of high-$Q$ microresonators without requiring any photodetection. Thermometric stabilization mitigates environmental as well as crosstalk-induced thermal perturbations, delivering a highly stable microresonator whose residual frequency fluctuations are only limited by $1/f$ noise in the control electronics. Realizing temperature-stabilized microresonators that are TRN-limited on long timescales is an exciting topic for future research, as it offers the prospect of precision metrology on a chip. We demonstrate that after a one-time calibration, the absolute resonance wavelength can be tuned to any desired value within the calibration range via thermometric control, with measured absolute-wavelength RMSEs $0.11$~pm and $<0.77$~pm after $12$~hours and $3$~days following the calibration, respectively. The emission wavelength of a DFB laser frequency-locked to the microresonator remains within $<\pm0.5$~pm of the mean over $50$ continuous hours, which is a $48\times$ improvement compared to free-running operation. This performance is superior to many commercial DFB and wavelength-locker-based laser systems. Finally, we stabilize a Kerr comb against strong thermal perturbations, offering a path toward chip-based comb devices that can potentially remain modelocked indefinitely in noisy environments.

Although our demonstration was focused on SiN microresonators, our technique is readily applicable to other dielectric and semiconductor photonic materials. Removing the requirement of drop-ports and photodetection for stabilization can significantly reduce design complexity and fabrication costs of practical photonic devices. Finally, the concepts described here can be applied to several other photonic integrated systems, including microresonator arrays and Mach-Zehnder interferometer (MZI) meshes for topological photonics~\cite{On2024} and photonic computing~\cite{Hamerly2022,Wu2023}.

\section{Acknowledgments}
The authors acknowledge support from the Army Research Office under Cooperative Agreement Number W911NF-24-2-0204 (M.L. and A.L.G.), the Air Force Office of Scientific Research under FA9550-20-1-0297 (M.L. and A.L.G.), and the Defense Advanced Research Projects Agency under HR0011-19-2-0014 (M.L. and A.L.G.).

The authors wish to acknowledge Kaylx Jang and Michael Cullen from the group of Prof. Keren Bergman for providing the photograph of a co-packaged silicon photonics device, previously reported in Ref.~\cite{Daudlin2025}, included in Fig.~\ref{fig1}. The authors thank Vivian Zhou from the group of Prof. Michal Lipson for assistance in setting up the vacuum system used in the temperature coefficient of resistance measurements.

\section{Author Contributions}
S.K.D. and A.L.G. conceived the project. S.K.D. and Y.Z. performed the experiments. S.K.D. and G.R.B. performed the design and characterization of the resistance thermometer. S.K.D., Y.Z., and A.L.G. performed the data analysis with input from all authors. K.J.M. fabricated the devices under the supervision of M.L. S.K.D. and A.L.G. wrote the manuscript with feedback from all authors. M.L. and A.L.G. supervised the project.

\section{Competing Interests}
S.K.D., Y.Z., K.J.M., G.R.B., M.L., and A.L.G. are inventors on a U.S. provisional patent application filed by The Trustees of Columbia University in the City of New York (application number 63/830,566) on 26 June 2025 that covers ultrastable microresonators using integrated thin-film resistance thermometers and thermistors.

\clearpage


\onecolumngrid
\setcounter{equation}{0}
\setcounter{figure}{0}
\setcounter{table}{0}
\setcounter{section}{0}
\renewcommand{\theequation}{E\arabic{equation}}
\renewcommand{\figurename}{Extended Data Figure}

\newpage

\begin{center}
\textbf{\large Frequency-stable nanophotonic microcavities via integrated thermometry: METHODS}
\end{center}
\section{Device Fabrication}
The on-chip resistance thermometer demonstrated in this work is similar to devices that have been extensively used as resistive microheaters~\cite{Miller:15,Joshi:16}. It consists of a Platinum thin-film resistor deposited on top of the oxide cladding layer. The resistors are fabricated by first patterning on top oxide layer, a stack of lift-off resist and positive photo-resist. Blanket Ti-Pt films are sputter deposited at room temperature, followed by metal lift-off to create resistors. A thin layer ($\approx 5\ \textrm{nm}$) of Titanium (Ti) is first deposited before deposition of Pt ($\approx 100\ \textrm{nm}$), to promote adhesion of Pt films. The sputtered metal adheres to and remains on the exposed oxide surface while metal on the photo-resist is stripped away in a resist stripping solution. The use of sputtered thin-films in our fabrication eliminates the need for processes requiring high-temperature and/or corrosive processing (e.g., electroplating).

\section{Calibrated Measurement of Absolute Resonance Frequency}

Main Text Figure~2(a) describes the experimental schematic for measuring the microresonator's absolute resonance wavelength. A weak ($\approx5$~$\mu$W) piezo-tuned laser source (Toptica CTL) is scanned across a single microcavity resonance at $200$~Hz, and the transmitted optical output is measured using a fast photodiode on a real-time oscilloscope. Calibration of the absolute resonance wavelength is performed as follows. A portion of the laser output is combined with a fixed-wavelength low-drift reference laser (Rio) and sent to a second fast photodiode. The generated heterodyne beat signal, measured using the same real-time oscilloscope, serves as a time marker for wavelength calibration. The wavelength of the reference laser is measured using a bench-top wavelength meter (Bristol Instruments 238A).

\section{Thermal Design for Integrated Thermometry}

\begin{figure*}[!htbp]
\centering
\includegraphics[width=\textwidth]{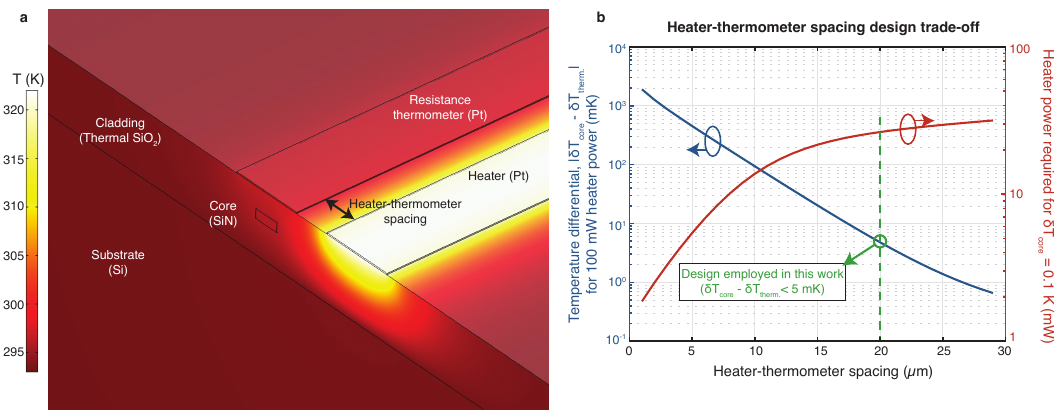}
\caption{\textbf{Thermal design for integrated thermometry.} \textbf{a}, Numerically simulated steady-state temperature profile for the scheme described in this work. The heater power is set to $100$~mW for the purposes of illustration. The initial temperature of the chip is $293$~K. \textbf{b}, Heater-thermometer spacing design trade-off, showing the core-thermometer temperature differential (left y-axis, in blue), and the heater power required for core heating of $0.1$~K (right y-axis, in dark red), versus heater-thermometer spacing.}\label{extDataFig:thermDesign}
\end{figure*}

Proper thermal design is essential for realizing an on-chip temperature-stabilization scheme that is effective at stabilizing the resonance frequency of the high-$Q$ microresonator. Specifically, the heater must be placed at a radial separation from the microresonator such that the any thermal correction it applies leads to equal temperature changes at the resistance thermometer as well as the microresonator core. The right heater-thermometer spacing is determined prior to device fabrication by solving the following heat transfer equation in steady-state using the COMSOL Multiphysics\textregistered~software:

\begin{equation}
    \rho C_p \frac{\partial T}{\partial t} + \nabla\cdot\boldsymbol{q} = Q+Q_{ted} 
\end{equation}

where $\rho$ is the material density, $C_p$ is the specific heat capacity at constant stress, $T$ represents the temperature, $q=-k\nabla T$ is the heat flux by conduction, $k$ is the thermal conductivity, $Q$ contains additional heat sources, and $Q_{ted}$ accounts for thermo-elastic damping in the material. All material thermal properties were set by the standard COMSOL material database for the cross-section of our devices. Figure~\ref{extDataFig:thermDesign}(a) displays the cross-section of our device, and the steady-state spatial temperature profile of the stationary thermal simulation. The heat power is set to $100$~mW for the purposes of illustration. The initial temperature of the entire chip, and the boundary condition temperature for the bottom face of the chip, are set to $293$~K. Here, we denote the resulting steady-state temperature change at the center of the SiN core and at the center of the resistance thermometer as $\delta$T\textsubscript{core} and $\delta$T\textsubscript{therm.}, respectively.

Figure~\ref{extDataFig:thermDesign}(b) shows the temperature differential $\delta$T\textsubscript{core} - $\delta$T\textsubscript{therm.} (left y-axis, in blue) as well as the heater power required to heat up the core (right y-axis, in dark red), which is a measure of heater efficiency, versus the heater-thermometer spacing. For small spacing, the heater-thermometer and heater-core separations are considerably different, leading to a large temperature differential. However, the close proximity allows for greater heater efficiency. For large spacing, the temperature differential drops, as the two separations are almost equal, while the heater efficiency worsens. The design employed in this work utilizes a $20$~$\mu$m heater-thermometer spacing, which leads to a very low temperature differential of $<5$~mK, making it effective in stabilizing the resonance frequency of the microcavity. The design principles described here can be readily applied to other integrated material platforms.


\section{Measurement of Temperature Coefficient of Resistance (TCR)}

\begin{figure*}[htbp]
\centering
\includegraphics[width=\textwidth]{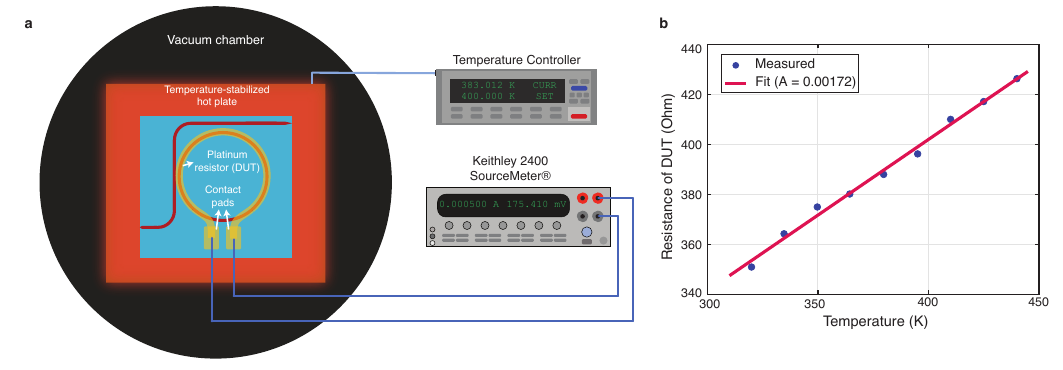}
\caption{\textbf{Measurement of temperature coefficient of resistance of the integrated Platinum thermometer.} \textbf{a}, Experimental schematic depicting a test chip, containing a platinum resistor (referred to here as device under test (DUT)) fabricated on top of a SiN microresonator, placed on a temperature-stabilized hot plate chuck inside a vacuum chamber. Its electrical resistance is probed as a function of temperature by performing current-voltage characterization using a Keithley $2400$ series SourceMeter\textregistered. \textbf{b}, Measured electrical resistance of DUT versus temperature (dark-blue filled circles) plotted alongside a linear fit (magenta solid line).}\label{extDataFig:TCR}
\end{figure*}

The integrated Platinum resistance thermometer, which allows the real-time measurement of the microresonator as described in the Main Text, is calibrated as shown in Fig.~\ref{extDataFig:TCR}(a). A test device is placed on a temperature-controlled chuck using a thermally-conductive adhesive in order to ensure proper thermal contact. The setup is placed and operated inside a vacuum probe station in order to minimize unwanted convective cooling. Two probes, connected to a Keithley $2400$ series SourceMeter\textregistered, are brought in electrical contact with the two ends of the Pt resistor under test (labeled as device under test (DUT) in Fig.~\ref{extDataFig:TCR}). At a given temperature of the chuck, the resistance $R$ of the DUT is measured by performing a current-voltage characterization. The voltage drop across the two contact pads is measured at various source currents, and a linear fit of the resulting data is obtained to extract $R$. This method is further explained in a prior manuscript~\cite{Bhatt2020}. The source current values used are low enough ($<1$~mA) to avoid any self-heating of the resistor. This process is repeated for a range of chuck temperatures $T$. Figure~\ref{extDataFig:TCR}(b) displays the results. Over the range of temperatures probed on this experiment, the resistance exhibits a linear trend. The temperature-coefficient of resistance is extracted by fitting the experimental data to Eq.~\ref{eq1}, and is determined to be $A=0.00172$. We note that DUT described here has a shorter fabricated length compared to the resistance thermometer described in the Main Text, leading to the lower base resistance.

\section{Measurement of Waveguide Thermo-Optic Coefficient}

A small-amplitude 2-Hz sinusoidal voltage is applied to the heater while monitoring the electrical resistance of the thermometer. Simultaneously, the shift in resonance frequency is probed using a scanning laser centered at 1560~nm, as described above. The free-spectral-range (FSR) of the SiN microresonator is 76 GHz, its loaded $Q$ is approximately $3\times10^{6}$, and the resonances in the telecommunications C-band are critically coupled. The width of the waveguide is 2100~nm. The measured change in thermometer resistance is converted to a change in temperature using the following equation,

\begin{equation}
    R(T)=R_0(1+A\Delta T)
    \label{eq1}
\end{equation}

where $R_0$ is the resistance at room temperature, $\Delta T$ is the change in temperature, and $A$ is the temperature coefficient of resistance (TCR). The coefficient $A$ is measured independently to be $A=0.00172$. The measurement procedure is detailed in Extended Data Fig. 2.

\clearpage

\section{Data Availability}
The data shown in this paper are available in this Zenodo repository: https://doi.org/10.5281/zenodo.17144471.

	\bibliography{therm.bib}


\onecolumngrid
\setcounter{equation}{0}
\setcounter{figure}{0}
\setcounter{table}{0}
\setcounter{section}{0}
\renewcommand{\theequation}{S\arabic{equation}}
\renewcommand{\thefigure}{S\arabic{figure}}

\newpage

\section{Low-noise Current Source}

\begin{figure*}[htbp]
\centering
\includegraphics[width=\textwidth]{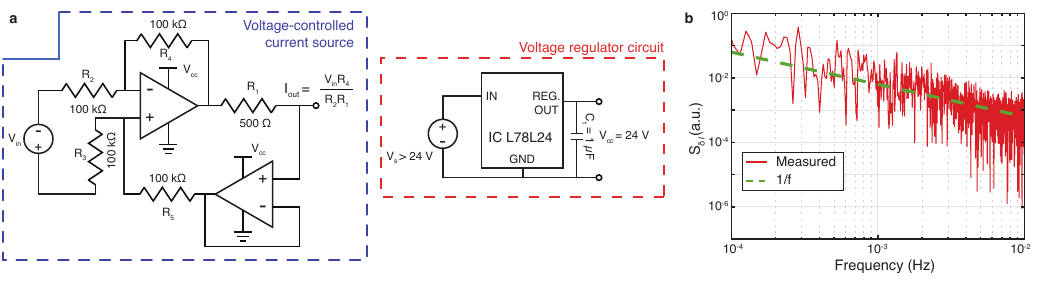}
\caption{\textbf{Design and performance characterization of low-noise current source.} \textbf{a}, Schematic of the custom-manufactured low-noise voltage-controlled current source employed in this work. Two opamps operate in a looped configuration, with the second opamp acting as a voltage follower. When resistors $R_{2-5}$ are equal, the gain is determined by $R_1$. A surface-mount voltage regulator with a $24$~V output is utilized to power the opamps in the precision current pump. \textbf{b}, Measured noise power spectral density (normalized) of the current source. It is observed to closely follow a $1/f$ trendline.}\label{supplFig:currSrc}
\end{figure*}

Figure~\ref{supplFig:currSrc}(a) depicts the schematic of the voltage-controlled current source (VCCS) used in this work. A two-operational-amplifiers  (opamps) topology that consists of one opamp feeding its outputs back to the other's positive differential input port was adopted. In steady state, when resistors $R_{2-5}$ are all chosen to be equal, the gain $I_{out}/V_{in}$ is simply given by $1/R_1$. A surface-mount voltage regulator with a low-noise $24$~V output powers the opamps and sources the current that drives the resistance thermometer. A printed circuit board (PCB) was designed and manufactured for this work. Figure~\ref{supplFig:currSrc}(b) shows the noise power spectral density (PSD) of the VCCS, measured as follows. The output current was sent through a low-noise bulk $559$~$\Omega$ resistor, and the voltage across it was measured using a real-time oscilloscope. Data from the oscilloscope was acquired onto a computer once every five seconds for a period of $24$~hours using a computer interface. The time series data of the voltage fluctuation across the bulk resistor was used to compute the noise power spectral density. It was assumed that the contribution of the bulk resistor to the measured noise is negligible. The resulting noise PSD closely follows a $1/f$ trendline for the frequencies of interest, demonstrating that the Allan Deviation of the stabilized microresonator is primarily limited by noise in the control electronics employed.

\section{Real-Time Thermometry during $24$-hour Measurements}

Figure~2(b) in the Main Text shows the measured resonance frequency shift for free-running ($\xi=0$), open-loop ($\xi=7.55$), and closed-loop ($\xi=7.55$) configurations over durations of $24$~hours each. Here, we demonstrate further that the voltage measured across the resistance thermometer yields the real-time temperature of the microresonator.

\begin{figure}[htbp]
\centering
\includegraphics{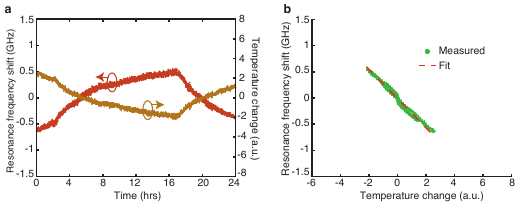}
\caption{\textbf{Real-time thermometry of free-running microresonator.} \textbf{a}, Measured cavity resonance shift (left y-axis) and temperature (right y-axis) for the free-running microresonator ($\xi=0$). The resonance frequency shift and temperature are observed to mirror each other, consistent with a positive thermo-optic coefficient. \textbf{b}, Scatter plot showing the linear relationship between the measured resonance frequency shift and temperature.}\label{supplFig:FR}
\end{figure}

Figure~\ref{supplFig:FR}(a) shows the measured resonance frequency shift for the free-running case on the left y-axis. The corresponding temperature measurement is shown on the right y-axis. The two quantities mirror each other's behavior versus time, consistent with a positive thermo-optic coefficient. Figure~\ref{supplFig:FR}(b) shows a scatter plot of the same data, with temperature change on the x-axis and the resonance frequency shift on the y-axis. The data agrees with the expected linear trend, which corresponds to a constant thermo-optic coefficient, which was measured and reported in Fig.~1 in the Main Text.

\begin{figure}[htbp]
\centering
\includegraphics{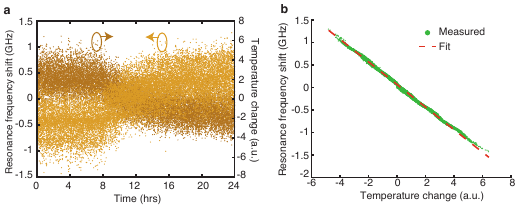}
\caption{\textbf{Real-time thermometry of open-loop microresonator.} \textbf{a}, Measured cavity resonance shift (left y-axis) and temperature (right y-axis) for the open-loop microresonator ($\xi=7.55$). The resonance frequency shift and temperature are observed to mirror each other, consistent with a positive thermo-optic coefficient. \textbf{b}, Scatter plot showing the linear relationship between the measured resonance frequency shift and temperature.}\label{supplFig:OL}
\end{figure}

Similarly, Fig.~\ref{supplFig:OL}(a) shows the measured resonance frequency shift for the open-loop case on the left y-axis. The corresponding temperature measurement is shown on the right y-axis. Once again, the two quantities mirror each other's behavior versus time. Figure~\ref{supplFig:OL}(b) shows a scatter plot of the same data, with temperature change on the x-axis and the resonance frequency shift on the y-axis. Because the open-loop case consists of additional perturbations injected via the heater ($\xi=7.55$), the microresonator experiences a larger magnitude of resonance frequency shift. As with the free-running case, the data agrees with the expected linear trend over the larger range.

\section{Comparison of Long-Term Stability with Packaged Commercial Laser Sources}
\begin{figure}[htbp]
\centering
\includegraphics{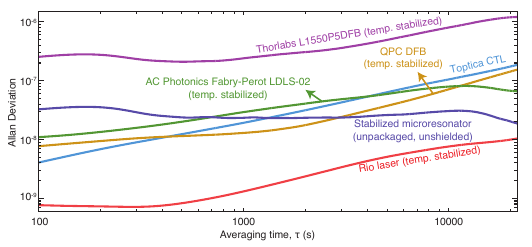}
\caption{\textbf{Comparison of long-term stability with packaged commercial lasers.} Measured Allan Deviation versus averaging time for the stabilized microresonator (unpackaged and unshielded) compared to several fully packaged, shielded, and temperature-stabilized commercial laser sources. The microresonator exhibits superior long-term stability and lower drift than most lasers tested during this study.}\label{supplFig:adevComp}
\end{figure}

In order to better place the performance of the stabilized microresonator in context, we compare the long-term in its absolute resonance wavelength with that of several commercial lasers. For the purpose of this comparison, our microresonator remains unshielded and unpackaged, whereas all of the lasers tested are fully packaged and shielded, and are temperature-stabilized with their dedicated commercial electronics. The procedure for measuring the microresonator's resonance wavelength is described in Fig.~2 in the Main Text. The emission wavelengths of the lasers are measured using a high-accuracy wavelength meter (Bristol Instruments 238A). Each measurement is carried out for 24 hours, and Allan Deviation is computed using the measured frequency data. The results are shown in Fig.~\ref{supplFig:adevComp}.

The stabilized microresonator exhibits superior long-term stability (i.e., lower Allan Deviation at longer averaging times) compared to all of the lasers included in the study with the exception of the Rio laser, the latter being a well-known industrial-grade laser system with low-noise and low-drift driving electronics. Further design optimization of the microresonator, such as increasing its mode volume and optimizing the position of the feedback heater, as well as improvements in the control electronics can greatly reduce the Allan Deviation of the microresonator.

\section{Soliton Microwave Carrier Frequency Stability}

Figure~6 in the Main Text describes the stabilization of a modelocked $\approx76.2$~GHz soliton comb generated in the microresonator. In Fig.~5\textbf{e}, we show the dramatic reduction in the frequency noise of the microwave generated via photodetection using a high-speed (100~GHz) photodiode. Here, we describe additional characterization of the microwave. A single-soliton Kerr comb is generated in the microresonator with a 300~mW CW pump power. The center frequency of the microwave carrier, i.e., the repetition rate of the soliton comb, is sampled once every five seconds using the phase noise analyzer setup shown in Fig.~6\textbf{a}. This measurement is carried out under two conditions, `open loop' and `stabilized'. In the open-loop case, the microresonator is subject to fluctuations in the ambient environment's temperature as well as noise introduced via the second heater ($\xi=0.75$). In the stabilized case, the microresonator remains in the same ambient environment, while the introduced white noise is dialed up to $\xi=7.55$.

\begin{figure}[htbp]
\centering
\includegraphics{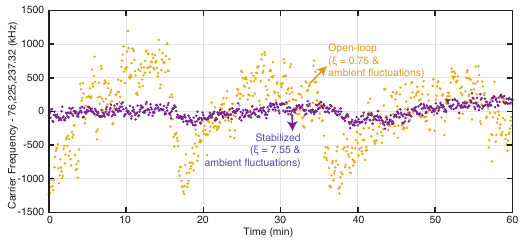}
\caption{\textbf{Microwave carrier frequency measurements.} Carrier frequency of the $\approx76.2$~GHz microwave generated via photodetection of the single-soliton comb, measured in the `open loop' ($\xi=0.75$) and `stabilized' ($\xi=7.55$) scenarios for one hour each. The stabilized case offers a more stable carrier frequency despite the significantly higher added noise.}\label{supplFig:muWaveCarrFreq}
\end{figure}

The injected noise level for the open-loop case is intentionally kept low enough ($\xi=0.75$) to allow the single-soliton Kerr comb to remain in its existence range throughout the measurement period. For stronger injected noise ($\xi\gtrsim1$), the soliton comb is pushed out if its existence range almost immediately. The open-loop data points in Fig.~\ref{supplFig:muWaveCarrFreq} reflect strong fluctuations on the time-scale of tens of minutes due to ambient fluctuations in the laboratory environment, as well as large shorter-term fluctuations resulting from the injected noise ($\xi=0.75$). When the feedback stabilization loop is turned on, the repetition rate of the single-soliton Kerr comb remains considerably more stable on shorter as well as longer timescales for significantly stronger injected noise ($\xi=7.55$), demonstrating that the method described here is effective at combating strong thermal crosstalk and other shorter-timescale fluctuations as well as strong ambient drift.

We note that due to the dependence of the repetition rate of the single-soliton comb to detuning, the resulting microwave carrier frequency is sensitive to changes in the intracavity optical power due to drift in the input optical coupling. Although we work with unpackaged devices throughout this manuscript, care was taken to ensure that the coupling drift throughout this measurement was minimal. The residual fluctuations in the stabilized case may be attributed to the impact of ambient temperature fluctuations on the source laser and the fiber amplifier used to generate the input pump, as the source laser was not locked to the microresonator during this experiment, and the pump intensity was not stabilized before coupling into the chip.



\end{document}